\documentclass[a4paper,12pt]{article}

\usepackage{amssymb}
\usepackage[numbers,sort&compress]{natbib}

\def\bea{\begin{eqnarray}}
\def\eea{\end{eqnarray}}
\def\be{\begin{equation}}
\def\ee{\end{equation}}
\def\ba{\begin{array}}
\def\ea{\end{array}}
\def\nn{\nonumber}

\newcommand{\bj}{\bar{\jmath}}

\newcommand{\bn}{\bar{n}}

\newcommand{\br}{\bar{r}}
\newcommand{\bs}{\bar{s}}
\newcommand{\bff}{\bar{f}}

\newcommand{\mV}{\mathcal V}

\newcommand{\dm}{\hat{d}}

\begin{document}
\baselineskip=16.6pt
\pagestyle{plain}

\renewcommand{\theequation}{\arabic{section}.\arabic{equation}}
\setcounter{page}{1}

\setlength\arraycolsep{2pt}

\begin{titlepage} 

\rightline{\footnotesize{DESY 08-038}} \vspace{-0.2cm}
\rightline{\footnotesize{CERN-PH-TH/2008-066}} \vspace{-0.2cm}
\rightline{\footnotesize{ZMP-HH/08-6}} \vspace{-0.2cm}

\begin{center}

\vskip 0.4 cm

{\LARGE  \bf de Sitter vacua in no-scale supergravities\\[2mm] and Calabi-Yau string models}

\vskip 0.7cm

{\large 
Laura Covi$^{a,b}$, Marta Gomez-Reino$^{c}$, Christian Gross$^{d}$, \\[1mm]
Jan Louis$^{d,e}$, Gonzalo A. Palma$^{a}$, Claudio A. Scrucca$^{f}$
}

\vskip 0.5cm

{\it
$^{a}$Theory Group, Deutsches Elektronen-Synchrotron DESY,\\ 
D-22603 Hamburg, Germany\\
$^{b}$Institute of Theoretical Physics, Warsaw University,
\mbox{ul. Hoza 69, 00-681 Warsaw, Poland}\\
$^{c}$Theory Division, Physics Department, CERN, \\
CH-1211 Geneva 23, Switzerland\\
$^{d}$II. Institut f\"ur Theoretische Physik, Universit\"at Hamburg, \\
D-22761 Hamburg, Germany\\
$^{e}$Zentrum f\"ur Mathematische Physik,
Universit\"at Hamburg,\\
Bundesstrasse 55, D-20146 Hamburg, Germany\\
$^{f}$Inst. de Th. des Ph\'en. Phys.,
Ecole Polytechnique F\'ed\'erale de Lausanne, \mbox{CH-1015 Lausanne, Switzerland}\\
}

\vskip 0.8cm

\end{center}

We perform a general analysis on the possibility of obtaining metastable vacua with spontaneously 
broken $\mathcal N=1$ supersymmetry and non-negative cosmological constant in the moduli 
sector of string models.
More specifically, we study the condition under which the scalar partners of the Goldstino are 
non-tachyonic, which depends only on the K\"ahler potential. This condition is not only 
necessary but also sufficient, in the sense that all of the other scalar fields can be given arbitrarily 
large positive square masses if the superpotential is suitably tuned. 
We consider both heterotic and orientifold string compactifications in the large-volume limit
and show that the no-scale property shared by these models severely restricts the allowed 
values for the `sGoldstino' masses in the superpotential parameter space. We find that a positive 
mass term may be achieved only for certain types of compactifications and specific Goldstino 
directions. Additionally, we show how subleading corrections to the K\"ahler potential which break 
the no-scale property may allow to lift these masses.

\end{titlepage}

\newpage

\section{Introduction} \setcounter{equation}{0}

It is widely believed that the existence of four-dimensional de Sitter (dS) vacua in low energy 
compactifications of string theory entails the presence of extended energy sources, such as 
D-branes, contributing to the vacuum energy density. This is motivated in part by the observation 
that smooth compactifications of \mbox{10-D} and \mbox{11-D} supergravities do not admit 
solutions to Einstein's equations characterised by both a positive cosmological constant and a 
stable ground state \cite{Gibbons, deWit, Maldacena:2000mw}.  
It has become clear, however, that this class of no-go theorems can be circumvented by 
including localised sources and/or taking into account higher order corrections in $\alpha'$ 
or the string coupling $g_{s}$ in the low energy analysis. In ref.~\cite{Giddings:2001yu} it was 
indeed shown that in type-IIB string theory compactified on Calabi-Yau orientifolds with
D-branes wrapping around cycles and nontrivial background fluxes a potential is generated 
for many of the scalar fields (moduli) present in the four-dimensional $\mathcal N=1$ supergravity.  
Including non-perturbative contributions all moduli can be stabilised but, generically, in a 
supersymmetric ground state which is either anti-de Sitter  or Minkowski \cite{Kachru:2003aw, 
Choi:2004sx, Balasubramanian:2005zx, lust, micu} 
whereas a positive cosmological constant necessarily requires 
the breaking of supersymmetry. For the `uplifting' from a supersymmetric vacuum to a dS 
vacuum a variety of mechanisms has been proposed and studied. For example, in 
ref.~\cite{Kachru:2003aw} it was shown that the joint contribution of non-perturbative effects
and an explicit supersymmetry-breaking term induced by anti-D3 branes can lead to a dS vacuum with 
fine-tuned cosmological constant and stable volume modulus.  Alternatively, there have been 
attempts to construct metastable vacua where supersymmetry is broken spontaneously either 
by $D$- or $F$-terms \cite{Burgess:2003ic, choi, vz, carlos, jeong, Saltman:2004sn, Becker:2004gw, 
Balasubramanian:2004uy, Saueressig:2005es, Intriligator:2006dd, Parameswaran:2006jh, 
Westphal:2006tn,crem, serone, Brax:2007xq, Dudas:2007nz, Silverstein:2007ac, Achu}.

Interestingly, there are no known examples of metastable vacua with spontaneously broken supersymmetry 
produced only by the volume moduli --or K\"ahler moduli-- in the absence of $\alpha'$ and worldsheet instanton 
corrections to the K\"ahler potential. At first sight this fact is a bit counter-intuitive. The superpotentials 
available in flux compactifications and/or compactifications on generalised \mbox{geometries} are 
sufficiently generic \cite{reviews} that one could expect no serious obstacle towards this end. 
Nevertheless, it was shown in ref.~\cite{Brustein:2004xn} that for $\mathcal N = 1$ supergravities
describing string compactifications with a single volume modulus $T$ and a no-scale K\"ahler potential  
\be
K = -3 \log (T + \bar T) \,,
\ee
stationary points of a positive scalar potential $V$ generated only by $F$-terms are always characterised 
by the existence of at least one tachyonic direction, independently of the superpotential $W=W(T)$. 
This result was made more precise in ref.~\cite{GomezReino:2006dk} and extended to more general situations,
and in particular to the class of compactifications in which the K\"ahler geometry spanned by the moduli is 
factorised into one or several sub-manifolds of constant curvature. More precisely, it was shown that also for 
the no-scale K\"ahler potential 
\be 
K = - \sum_{i} n_{i} \log (T^{i} + \bar T^{i}), \qquad \mathrm{with} \qquad  \sum_{i} n_{i} = 3, \label{K-fact}
\ee
stationary points of a positive scalar potential $V$ have at least one tachyonic direction independently of
$W$. Moreover, this tachyonic direction was shown to become marginally flat only when the superpotential 
$W$ is chosen in such a way that $V=0$. Similar results were derived in ref.~\cite{GomezReino:2006wv} for coset manifolds arising in orbifold compactifications. This new class of no-go theorems --which relies only on the 
properties of the K\"ahler potential-- raises the natural question about the role of the volume moduli in the 
construction of metastable vacua in more generic string compactifications where the K\"ahler geometry 
spanned by the moduli becomes nontrivial.

The purpose of this paper is twofold. First we refine the previous analysis of four-dimensional $\mathcal N = 1$ 
supergravities given in refs.~\cite{GomezReino:2006dk, GomezReino:2006wv,GomezReino:2007qi} by 
emphasising that the crucial quantity to study in order to achieve vacuum metastability is the mass of the scalar
superpartners of the Goldstino. We show that all of the other scalar fields can be made arbitrarily massive by 
appropriately choosing the superpotential. However, this is not the case for the two sGoldstinos since in the 
limit of global supersymmetry the Goldstino is exactly massless and therefore the sGoldstinos can never get a 
mass from the superpotential. Instead their masses are generated by the supersymmetry breaking mechanism 
with their mass-difference being of the order of the supersymmetry breaking scale. As a consequence their masses 
are not necessarily positive. It is precisely this fact which is at the heart of the problem of identifying locally stable dS 
vacua. From this discussion it is also immediately clear that a positive sGoldstino mass is a necessary condition 
for the metastability of any dS vacua and, furthermore, this condition does not depend on the superpotential 
but only on the form of the K\"ahler potential. This observation considerably simplifies the search for a viable 
dS ground state.

The second aspect of this paper concerns an analytical study of specific classes of $\mathcal N = 1$
supergravities which appear as the low energy limit of string compactifications. We show that there 
exist entire classes of compactifications which do not admit any metastable dS vacua, irrespectively of 
the superpotential or the vacuum expectation values that the moduli may acquire.  For instance, we show 
that de Sitter vacua are excluded in the case of $K3$ fibrations regardless of the number of moduli or their 
vacuum expectation values. On the other hand, we also identify particular classes of compactifications in 
which the necessary conditions are indeed fulfilled and thus viable dS vacua should exist. Let us stress 
here that we do not minimise any explicit potential. Rather we study the condition for the existence of 
dS vacua and show that irrespectively of the superpotential this condition is not easily satisfied. 
We think that this is the reason for the difficulties encountered in constructing explicit metastable de Sitter 
vacua in low energy compactifications of string theory.

The organisation of this paper is as follows. In Section \ref{sugra}, we start by reviewing the conditions under 
which a generic  supergravity model with chiral multiplets admits viable vacua with spontaneously broken 
supersymmetry and non-negative cosmological constant. Then in Section  \ref{S:LVS} we apply the resulting 
condition to the 
class of models where the K\"ahler potential satisfies either the no-scale property or a more restrictive homogeneity property respected by  large-volume scenarios of string theory. In Sections \ref{heterotic} and \ref{orientifolds} 
we study the large-volume limit  of  heterotic and orientifold models respectively and derive in each case the 
form of the metastability condition. There we also apply our general results to classes of models where the 
metastability condition can be studied analytically and show explicitly that a positive square mass may be 
achieved only for certain types of compactifications and particular Goldstino directions. We also study the 
effect of (subleading) $\alpha'$ corrections to the K\"ahler potential and show that they contribute to the 
sGoldstino masses and can render them positive even for those models where it is not possible at leading order.  
Finally, in Section \ref{conclusions} we present our conclusions.

\section{Metastable vacua in supergravity}  \label{sugra}
\setcounter{equation}{0}

In this section, we briefly review and extend the strategy that was presented in
\mbox{refs.~\cite{GomezReino:2006dk, GomezReino:2006wv,GomezReino:2007qi}} to 
study the stability of non-supersymmetric vacua in general supergravity models with $\mathcal N = 1$ 
supersymmetry in four dimensions.\footnote{A similar strategy has also been used in ref.~\cite{DD}
to explore the statistics of supersymmetry breaking vacua in
certain classes of string models.}
We assume that vector multiplets play a negligible role in the dynamics of supersymmetry 
breaking, and focus thus on theories with only chiral multiplets.

Recall first that the most general two-derivative Lagrangian for a supergravity theory with $n$ chiral superfields 
is entirely defined by a single arbitrary real function $G$ depending on the corresponding chiral superfields $\Phi^i$ 
and their conjugates $\bar \Phi^{\bar \imath}$. Derivatives with respect to $\Phi^i$ and $\bar \Phi^{\bar \jmath}$ are 
denoted by lower indices $i$ and $\bar \jmath$. Using Planck units where $M_P=1$, the function $G$ can be 
decomposed in terms of a real K\"ahler potential $K$ and a holomorphic superpotential $W$ in the following 
way:
\be
G(\Phi,\bar \Phi) = K(\Phi,\bar \Phi) + \log W(\Phi) + \log \bar W(\bar \Phi) \,.
\ee
The quantities $K$ and $W$ are however defined only up to K\"ahler transformations acting as
$K \to K + f + \bar f$ and $W \to W e^{-f}$, where $f$ is an arbitrary holomorphic function of the superfields. 
The bosonic part of the action takes the form:
\begin{eqnarray}
S =  \int \!\! \sqrt{-g} \, \Big[ \frac{1}{2}  R - g_{i \bar \jmath} \, \partial \phi^{i} \partial \bar \phi^{\bar \jmath} 
- V(\phi,\bar \phi)  \Big] \,.
\label{s1: act}
\end{eqnarray}
The K\"ahler metric $g_{i \bar \jmath} = K_{i \bar \jmath} = \partial_{i} \partial_{\bar \jmath} K$ is used to raise and lower indices, and defines a K\"ahler 
geometry for the manifold spanned by the scalar fields. It is assumed to be positive definite, such that the 
scalar's kinetic energy is positive. The potential takes the following simple form:
\begin{eqnarray}
V = e^{G} \big(G^{i} G_{i} - 3  \big)\,. 
\label{s2:F-term-G} 
\end{eqnarray}
The auxiliary fields of the chiral multiplets are fixed by their equations of motion to be $F^i=m_{3/2}G^i$
with a scale set by the gravitino mass $m_{3/2}=e^{G/2}$. Whenever $F^i\neq 0$ on the vacuum, 
supersymmetry is spontaneously broken, and the direction $G^i$ in the space of chiral fermions 
defines the Goldstino which is absorbed by the gravitino in the process of supersymmetry breaking.

\subsection{Condition for metastability} \label{stability analysis} 

Supersymmetry-breaking metastable vacua with non-negative cosmological constant are associated to local 
minima of the potential at which $F^i \neq 0$ and $V \ge 0$. These vacua can be classified by looking 
at stationary points with $V'=0$, imposing that the value of the potential should not be negative, $V\ge 0$,
and finally requiring that the Hessian matrix should be positive definite: $V''>0$. 

The derivatives of the potential (\ref{s2:F-term-G}) are most conveniently computed by using the 
covariant derivative $\nabla_{i}$ defined by the K\"ahler metric $g_{i \bar \jmath}$, and the associated 
Riemann curvature tensor $R_{i \bar \jmath m \bar n}$. The first derivative is just $V_i = \nabla_i V$, 
and the stationarity conditions $V_i = 0$ read
\bea
e^{{G}} \left({G}_{i} + {G}^{k} \nabla_{i} {G}_{k}\right) + G_i V = 0\,.
\label{stationary1}
\eea
The second derivatives of the potential can also be computed by using covariant derivatives, 
since the extra connection terms vanish by the stationarity conditions. There are two different 
$n$-dimensional blocks, $V_{i \bar \jmath} = \nabla_i \nabla_{\bar \jmath} V$ and
$V_{i j} = \nabla_i \nabla_j V$, and these are found to be given by the following expressions:\footnote{Our 
conventions for the Riemann tensor are given by eq.~(\ref{eq:generalRiemann}) in the Appendix.}
\begin{eqnarray}
V_{i \bar \jmath} &=&  e^{G} \left(G_{i \bar \jmath}  +  \nabla_i G_k \nabla_{\bar \jmath}  G^k
-  R_{i \bar \jmath m \bar n} \, G^m  G^{\bar n} \right) + \left(G_{i \bar \jmath}  -  G_i G_{\bar \jmath}\right) V \,,
\label{mass1} \\[1mm]
V_{i j} &=& e^{G} \left(2 \nabla_i G_j+  G^k \nabla_i \! \nabla_j G_k \right)
+ \left(\nabla_i G_j -  G_i G_j \right) V \,.  
\label{mass2}
\end{eqnarray}
The metastability condition is then the requirement that the whole $2n$-dimensional Hessian mass 
matrix $M^2$ should be positive definite, where
\begin{eqnarray}
M^2 = \left(\matrix{
V_{i \bar \jmath} & V_{i j} \nn \\
V_{\bar \imath \bar \jmath} & V_{\bar \imath j}} 
\hspace{-35pt}
\right) \,.
\end{eqnarray}

It is clear that for a fixed K\"ahler potential $K$, most of the eigenvalues of this mass matrix 
can be made positive and arbitrarily large by suitably tuning the superpotential $W$. More precisely,
the $n-1$ chiral multiplets that are orthogonal to the Goldstino multiplet can acquire a large overall 
supersymmetric mass contribution from $W$, which can overcome the mass splitting of 
order $m_{3/2}$ induced by supersymmetry breaking, and lead to positive square masses for the 
scalar field components. The Goldstino multiplet, on the other hand, cannot receive any supersymmetric 
mass contribution from $W$, since in the limit of rigid supersymmetry its fermionic 
component must be massless. The mass splitting of order $m_{3/2}$ induced by supersymmetry 
breaking can then potentially make the square mass of the scalar field component negative. 

From a more technical point of view, this conclusion can be obtained by recalling that derivatives 
of $G$ with mixed holomorphic and antiholomorphic indices depend only on $K$, while quantities 
like $G_i$, $\nabla_{i} G_{j}$ and $\nabla_{i} \nabla_{j} G_{k}$ depend also on $W$, and more
precisely on $(\log W)_i$, $(\log W)_{ij}$ and $(\log W)_{ijk}$. Keeping $K$ fixed and tuning $W$, 
one can then vary in an arbitrary way these quantities. This allows to adjust first the quantities 
$\nabla_i \nabla_j G_k$ to set the block $V_{i j} $ to zero, and next the quantities $\nabla_i G_j$ 
to make most of the eigenvalues of $V_{i \bar \jmath}$ positive. On top of that, one still has the freedom 
of arbitrarily choosing $G_i$. The only restriction in the second step comes from the fact that the 
projection of $V_{i \bar \jmath}$ along the Goldstino direction $G^i$ is actually fixed by the stationarity 
condition (\ref{stationary1}), and can therefore not be adjusted. This means that the square masses of 
the two sGoldstinos cannot be arbitrarily shifted by adjusting $W$, and that their value crucially 
depends on $K$. 

In order to study metastability, it is thus sufficient to study the projection of the diagonal block 
$V_{i \bar \jmath} $ of the mass matrix along the Goldstino direction $G^i$. More precisely, 
we find it convenient to rescale this quantity by the overall mass scale $m_{3/2}^2$ and 
consider the following parameter:
\begin{eqnarray}
\lambda = e^{-G} \,  V_{i \bar \jmath} \, G^i G^{\bar \jmath} \,.
\end{eqnarray}
Strictly speaking $\lambda$ is a linear combination of eigenvalues of $V_{i \bar \jmath}$ 
with non-negative coefficients in front of them. It therefore defines a natural mass scale 
$\tilde m^{2} \equiv e^{G} \lambda / G^{i} G_{i}$ which can be thought of as the mass obtained 
by projecting $V_{i \bar \jmath}$ along the Goldstino direction $G^{i}$. Accordingly, we identify
here $\tilde m$ with the mass of the sGoldstinos.

By using eqs.~(\ref{stationary1}) and (\ref{mass1}), one can compute $\lambda$ more explicitly.
The result is found to depend only on the parameters $G^i = e^{-G/2} F^i$ defining the direction 
of supersymmetry breaking, contracted with the metric and the Riemann tensor of the scalar geometry:
\begin{eqnarray}
\lambda = 2 \, g_{i \bar \jmath} \, G^i  G^{\bar \jmath}  
-  R_{i \bar \jmath m \bar n} \, G^{i}  G^{\bar \jmath} G^{m} \! G^{\bar n}  \,. \label{lambda1}
\end{eqnarray}
For given $K$ and arbitrary $W$, the quantities $G^i$ can be varied but the metric and the Riemann
tensor are fixed. One can then look for the preferred direction that maximises $\lambda$.\footnote{See 
ref.~\cite{Lukas} for an algebraic method for finding the minima for a wide class of superpotentials.}
If $\lambda_{\rm max} < 0$, then one of the sGoldstinos is unavoidably tachyonic, and the vacuum is 
unstable. If instead $\lambda_{\rm max} >0$, then the sGoldstinos can be kept non-tachyonic 
by choosing $W$ such that the Goldstino direction is close enough to the preferred direction, and more
precisely inside a cone for which $\lambda \in [0,\lambda_{\rm max}]$. As already mentioned, the rest of 
the scalars can always be given a positive square mass by further tuning $W$.
The crucial condition for metastability, which constrains both the K\"ahler geometry and the supersymmetry
breaking direction, is then \cite{GomezReino:2006dk}
\be
\lambda > 0 \,.
\ee

\subsection{Analysis of the metastability condition} \label{S:analysis} 

The implications of the metastability condition $\lambda > 0$ have been studied in 
refs.~\cite{GomezReino:2006dk, GomezReino:2006wv} for models with a fixed cosmological 
constant. But one can actually perform a similar study without specifying the value of the cosmological 
constant and only requiring that it is non-negative. It is clear from the form of eq.~(\ref{lambda1}) 
that for sufficiently small values of the $G^{i}$, it would always be possible to find configurations 
such that $\lambda > 0$, since the quartic term becomes subdominant and the quadratic term is positive. 
However, in this regime the cosmological constant would necessarily be negative. 
Whenever some of the $G^i$ are instead of order $1$, as required to achieve a non-negative 
cosmological constant, the quadratic and quartic terms compete, and the existence of configurations
with $\lambda > 0$ strongly depends on the form of the curvature tensor. To analyse the rather constrained 
problem of finding whether there exist vacua with $V \ge 0$ and $\lambda > 0$ it is convenient to rewrite 
$\lambda$ as the sum of two pieces,
\begin{eqnarray}
\lambda =  - \frac{2}{3} e^{-G} V \big( e^{-G} V  + 3 \big) + \sigma  
\label{lambda},
\end{eqnarray}
where $\sigma$ is defined to be
\begin{eqnarray}
\sigma = \Big[\frac{1}{3} \left(g_{i \bar \jmath} \, g_{m \bar n} + g_{i \bar n} \, g_{m \bar \jmath} \right)
- R_{i \bar \jmath m \bar n} \Big] G^{i} G^{\bar \jmath} G^{m} \! G^{\bar n} \,. 
\label{sigma}
\end{eqnarray}
As long as $V>0$ the first term in eq.~(\ref{lambda}) is always negative and its precise value depends only on the length of 
the vector $G^i$ which determines the cosmological constant. The second term 
in eq.~(\ref{lambda}) has instead a sign that depends only on the orientation of the vector $G^i$, and not on 
its length. Therefore, the possibility of finding solutions to the metastability condition $\lambda > 0$
depends exclusively on the sign of $\sigma$. Indeed, starting from any $G^i$ such that $\sigma (G^i) > 0$, 
one can always tune the superpotential $W$ to rescale $G^i$ by some real factor $r$ to achieve $V(r G^i) = 0$ and thus $\lambda(r G^i) > 0$,
proving the existence of Minkowski vacua. Moreover, by slightly increasing $r$ one can make $V(rG^i) > 0$ 
and still keep $\lambda(r G^i) > 0$, achieving thereby de Sitter vacua. For a fixed value of the gravitino mass scale $m_{3/2} = e^{G/2}$ it is however clear that how big a 
cosmological constant $V$ can be achieved while keeping $\lambda > 0$ depends on the size of $\sigma$ 
for the reference situation where $V(G^i) = 0$. The same kind of reasoning tells us that if $\sigma < 0$ 
for all the possible orientations of $G^i$, then one can never achieve $V \ge 0$ and $\lambda > 0$ 
simultaneously.
We can therefore conclude that the analysis of the sign of the function $\lambda$ for non-supersymmetric 
vacua with $V \ge 0$ is equivalent to the analysis of the sign of the function $\sigma$ without specifying the
value of the cosmological constant. More 
precisely, the condition for the existence of viable vacua is that
\bea
\sigma > 0 \,.
\eea

It is easy now to check a few well known results concerning the existence of metastable vacua. 
Consider for instance those models where the K\"ahler potential is of the canonical form 
$K = \sum_{i} |\Phi^{i}|^{2}$ for which the K\"ahler manifold has a vanishing Riemann tensor. 
In this case one has
\be
\sigma = \frac{2}{3} (G^{i} \bar G_{i})^{2} > 0 \,,
\ee
and no obstruction is met towards the construction of metastable vacua. Another simple example is provided by 
string compactifications described by a single volume modulus $T$ and a no-scale K\"ahler potential of the 
form $K = - 3 \log (T + \bar T)$. In this case, one finds that
\be
\sigma = 0 \, ,
\ee
independently of the value $G^{T}$, and thus dS vacua are excluded \cite{Brustein:2004xn} (see also \cite{BenDayan}). 
Finally, models with separable $K = - 3 \log (T + \bar T) + \sum_{i} |\Phi^{i}|^{2}$ also grant the existence of de Sitter 
vacua as long as $G^i \neq 0$. If $W$ is separable as well, so that the $2$ sectors interact only gravitationally, it is 
actually possible to uplift any would-be supersymmetric minimum in the $T$ sector with a $\Phi^i$ sector breaking 
spontaneously supersymmetry well below the Planck scale \cite{GomezReino:2006dk}. See  \cite{Lebedev:2006qq} 
for a generalization to a certain class of non-separable $W$, and \cite{Fup,dudas} for specific examples. 
On the other hand, for similar models with non-separable $K = - 3 \log (T + \bar T - 1/3 \sum_{i} |\Phi^{i}|^{2})$, as 
those considered in ref.~\cite{LS}, the scalar manifold is maximally symmetric and one finds again 
$\sigma = 0$ \cite{GomezReino:2006wv}. See ref.~\cite{Achu} for a recent general study of this type of uplifting.

Notice that $\sigma$ has the very useful property of being a homogeneous function of degree $(2,2)$ 
in the variables $(G_{i},  G_{\bar \jmath})$, meaning that
\be
G_i\frac{\partial \sigma}{\partial G_i}=G_{\bar \jmath}\frac{\partial \sigma}{\partial G_{\bar \jmath}}=2\sigma \,.
\ee 
As a consequence of this property, any stationary point of $\sigma$ as a function of $G_{i}$ 
leads to $\sigma=0$. This implies in turn that, at any given point in the K\"ahler manifold spanned by the chiral fields, 
the function $\sigma$ can have only one such 
stationary point, or a degenerate family of them, with $\sigma = 0$. This is due to the fact that if the 
value of the function becomes non-zero when moving away from such a stationary point, then 
its first derivative is no longer allowed to vanish again.

Based on this property, it is possible to outline a general and systematic procedure to find out whether $\sigma > 0$ 
can be achieved in a particular model by only requiring that the set of points $G_i^0$, at which $\sigma$ becomes 
stationary, is known.  Indeed, it is sufficient to study the convexity of the function $\sigma(G^{i})$ in the vicinity of 
$G_i^0$ by scanning  all the orientations of $G_i$ away from $G_i^0$ for which $\sigma$ is allowed to grow. 
If $\sigma(G_i^0) = 0$ is a local minimum then, by the method described before, any direction $G_i \neq G_i^0$ 
may be rescaled to render a metastable vacuum. If instead $\sigma(G_i^0) = 0$ turns out to be a maximum, then 
one is forced to exclude the K\"ahler potential $K$ of the model as a possible candidate to generate metastable 
vacua. Finally, if $\sigma(G_i^0) = 0$ turns out to be a saddle point, then only a reduced subset of orientations 
$G_{i}$ will qualify to render metastable vacua. We should bear in mind, however, that the metric and the Riemann
tensor appearing in the definition of $\sigma$ depend on the values of the scalar fields. Therefore, one should 
also scan over the allowed values of $\phi^{i}$.

The  procedure just described is very useful and in principle simple to implement when the convexity of the function 
$\sigma$ cannot be determined analytically. This is particularly the case of the class for models appearing in large 
volume compactifications of string theory. As we show in the next section, the scaling properties respected by the type 
of K\"ahler potentials appearing in such scenarios imply two important properties of the function $\sigma$:
first, stationary points of $\sigma$ are of the form $G^i \propto K^i$, and second, such points are either of the 
saddle-point type or maxima. One is then left with the task of determining, by studying the vicinity of $G^i \propto K^i$, 
which one of these two situations is being dealt with. 

\section{Metastability in large-volume scenarios} \label{S:LVS} 
\setcounter{equation}{0}

We now focus on some generic properties respected by models emerging 
in large-volume scenarios of string theory.
More specifically, we apply the analysis of the previous section to the class of models where the 
K\"ahler potential satisfies either the no-scale property or an even more restrictive scaling property. 

\subsection{No-scale models} \label{S:no-scale} 

A common characteristic found in string compactifications is the no-scale property \cite{Cremmer}
\begin{eqnarray}
K^{i}K_{i} = 3 \,,\label{nosc}
\end{eqnarray}
which holds for the K\"ahler moduli parameterising the shape and size of the 
compactified volume in the large-volume limit. 
Similarly, it also holds for the complex structure moduli in the large-complex-structure limit.
We would then like to study the function $\sigma$ as defined in (\ref{sigma}) for the particularly relevant 
class of supergravity models satisfying this no-scale property, in order to understand whether this restriction  
implies any useful information concerning metastability.

The simplest examples of such no-scale models are certain coset manifolds of the type 
$SU(p,q)/(U(1)\times SU(p) \times SU(q))$ and $SO(2,2+p)/(SO(2) \times SO(2+p))$, with 
appropriate constant curvature, arising in orbifold string models. Due to the fact that they are 
homogeneous and symmetric, these particular spaces lead to a simple form of the Riemann tensor. 
The implications of the stability condition can then be worked out completely. It was in fact shown 
in \cite{GomezReino:2006wv} that in these models the maximal value of $\sigma$ is precisely 
zero, and that this value is obtained for the particular direction $G^i = K^i$, or equivalent directions 
related to this by the isometries of the space.

In more complicated situations where the curvature is not constant, like in Calabi-Yau models with 
and without orientifolds, the Riemann tensor takes a more complicated form and the study of the metastability 
condition becomes substantially more complicated. However, since the property (\ref{nosc}) is 
valid at any point of the K\"ahler manifold, it implies some simple and nontrivial restrictions 
on the Riemann tensor, and in particular on its contractions with the special vector $K^i$. 
For instance, taking one derivative of (\ref{nosc}) one finds
\begin{eqnarray}
K_i + K^k \nabla_i K_k = 0,
\end{eqnarray}
whereas taking two derivatives one deduces the following relations:
\begin{eqnarray}
&& g_{i \bar \jmath} + \nabla_i K_k  \nabla_{\bar \jmath}  K^k - R_{i \bar \jmath m \bar n}  K^m \! K^{\bar n} = 0\,, \\
&& 2\, \nabla_i K_j +  K^k \nabla_i \nabla_j K_k = 0\,.
\end{eqnarray}
Contracting the first of these relations with $K^i K^{\bar \jmath}$ and $K^{\bar \jmath}$ respectively, one can
then derive the relations
\begin{eqnarray}
&& R_{i \bar \jmath m \bar n} K^{i} K^{\bar \jmath} K^{m} \! K^{\bar n} = 6\,,  \label{RKKK} \\ 
&& R_{i \bar \jmath m \bar n} K^{\bar \jmath} K^m \! K^{\bar n} = 2 K_i \,.     \label{RKKKK}
\end{eqnarray}

These relations are useful to study the function $\sigma$ for this class of models. In order to do so, it is natural to 
introduce the projector onto the subspace orthogonal to $K^i$, since we know that at least in the particular case 
of constant curvature manifolds this is the special direction that maximises $\sigma$. Thanks to the no-scale 
property, this projector is simply
\be
P_i^j = \delta_i^j - \frac 13 K_i K^j \,.
\ee
We can then decompose the vector $G_i$ into two independent pieces, one parallel to $K_i$ and parameterised 
by a numerical coefficient $\alpha$, and one orthogonal to $K_i$ and parameterised by a vector $N_i$ satisfying 
$N^i K_i = 0$:
\be
G_i = \alpha K_i + N_i \,.
\label{decomp}
\ee
The quantities $\alpha$ and $N^i$ are given by
\begin{eqnarray}
N_{i} = P_i^j G_j \,, \qquad 
\alpha = \frac13 K^iG_i \,. \label{nialfa}
\end{eqnarray}
The function $\sigma$, as defined in eq.~(\ref{sigma}), may then be expressed in terms of the independent 
quantities $\alpha$ and $N_{i}$ in the following way:
\begin{eqnarray}
\sigma &=&  4 |\alpha|^{2} \left(g_{i \bar \jmath} - R_{i \bar \jmath m \bar n} K^m \! K^{\bar n} \right) N^i N^{\bar \jmath} 
- \left(\bar \alpha^{2} R_{i \bar \jmath m \bar n} K^i K^m \! N^{\bar \jmath} N^{\bar n} + {\rm c.c}\right) \nn \\[2mm]
&& - {2}\, \left(\bar\alpha R_{m \bar n i \bar \jmath} K^{m} \! N^{\bar n} N^{i}  N^{\bar \jmath} + {\rm c.c} \right) \nn \\
&& + \Big[\frac{1}{3} \left(g_{i \bar \jmath} \, g_{m \bar n} + g_{i \bar n} g_{m \bar \jmath} \right)
- R_{m \bar n i \bar \jmath} \Big] N^{i}  N^{\bar \jmath} N^{m} \! N^{\bar n} \,.
\end{eqnarray}
Note that this result is at least quadratic in the variables $N^i$. This implies that there is a degenerate 
family of stationary points for $N^i = 0$ and arbitrary $\alpha$, that is for $G^i \propto K^i$, with value 
$\sigma = 0$. To say more about the convexity of $\sigma$ at this set of points we still require some more information
regarding contractions between $K_{i}$ and the Riemann tensor. As we will see in the following, this additional information can be obtained by imposing an extra condition generically respected by large-volume string compactifications.

\subsection{Real homogeneous no-scale models}   \label{S:homogeneous no-scale}

A more restrictive property characterising large-volume scenarios is that their K\"ahler potential depends only on the real part of the superfields and exhibits therefore $n$ independent shift symmetries, under which 
$\delta_i \Phi^j = i \epsilon \delta_i^j$ with constant $\epsilon$. This means in particular that any distinction between 
holomorphic and antiholomorphic indices can be dropped. Furthermore, it turns out that there 
exists a coordinate frame where $e^{-K}$ is a homogeneous function of degree $3$ in the fields 
$\Phi^i + \bar \Phi^i$. This implies that
\be
- \big(\Phi^i + \bar \Phi^i \big) K_i = 3 \,. \label{hom}
\ee
Taking a derivative, it then follows that
\be
K^i = - \big(\Phi^{i} + \bar \Phi^{i} \big)  \,.
\ee
This equation guarantees, together with the previous one, that the no-scale property $K^i K_i = 3$ is satisfied.
But taking a derivative, it also implies that $\partial_i K^j = - \delta_i^j$, which after lowering the indices implies
\be
K_{i j m} K^m = 2\, g_{ij} \,.
\ee
Taking another derivative of this, one finds also
\be
K_{i j m n} K^m = 3 K_{i j n} \,.
\ee
From these two equations, it follows then that
\bea
R_{i j m n} K^{m} &=& K_{i j n} \,, \label{RK} \\
R_{i j m n} K^m \! K^n &=& R_{i m j n} K^m \! K^n = 2\, g_{ij} \,. \label{RKK}
\eea
Finally, contracting these equations with one and two more $K^k$'s and using the no-scale condition, 
one also recovers the same relations (\ref{RKKK}) and  (\ref{RKKKK}) holding for general no-scale models.

It is convenient at this point to introduce a new 
notation to deal with complex quantities such as $G_{i}$ and $G_{\bar \imath}$ in such a way that the 
bar does not appear on top of the indices. Compared to the usual notation, we introduce the following 
substitutions: $G_i \to G_i$, $G_{\bar \imath} \to \bar G_i$, $G^i \to \bar G^i$, $G^{\bar \imath} \to G^i$. 
Similarly, for the $N_{i}$'s we use:  $N_i \to N_i$, $N_{\bar \imath} \to \bar N_i$, $N^i \to \bar N^i$, 
$N^{\bar \imath} \to N^i$.

Using eqs.~(\ref{RKKK}), (\ref{RKKKK}), (\ref{RK}) and (\ref{RKK}), and decomposing as before 
$G_i = \alpha K_i + N_i$ and $\bar G^i = \bar \alpha K^i + \bar N^i$, one finds that the function 
$\sigma$ takes in this case the following form:
\begin{eqnarray}
\sigma &=& - 2\, \left( \alpha \bar N^i + \bar \alpha  N^i \right)
\left(\alpha  \bar N_i + \bar \alpha  N_i \right)
- {2}\, K_{i m n}  \left(\alpha \bar N^i + \bar\alpha N^i \right) N^m \! \bar N^n \nn \\
&& + \Big[\frac{1}{3} \left(g_{i j}\, g_{m n} + g_{i n} \, g_{m j} \right)
- R_{i j m n} \Big] N^i \bar N^j N^m \! \bar N^n  \,.
\label{sigmaalphaN}
\end{eqnarray}
This result shows that $\sigma$ has a local maximum with value $0$ at $N_i = 0$ at quadratic order in the 
$N^i$ variables for orientations of $G^i$ characterised by 
$\alpha \bar N^i + \bar \alpha  N^i \neq 0$. Nevertheless, this does not imply that $\sigma$ is negative 
definite, because when $\alpha \bar N^i + \bar \alpha  N^i = 0$ the potential 
is flat at the quadratic and cubic orders and its convexity is determined by the quartic terms in $N_i$. In order
to gain further insight it is useful to complete the squares in the variable 
$\alpha \bar N^i + \bar \alpha  N^i$ and rewrite $\sigma$ in the form
\be
\sigma = - 2 \, s^i s_i +  \omega  \,,
\label{sigmaomegas}
\ee
where
\begin{eqnarray}
s^i &=& \alpha \bar N^i + \bar \alpha  N^i + \frac 12 P^{ij} K_{j m n} N^m \! \bar N^n \, , \label{sn} \\
\omega &= & \Big[\frac{1}{3} \big(g_{i j} \, g_{m n} + g_{i n} \, g_{m j} \big)
- R_{i j m n} + \frac 12 K_{i j k} P^{k l} K_{l m n}  \Big] N^i \bar N^j N^m \! \bar N^n  \, .
\label{omega} 
\end{eqnarray}
Observe now that all the dependence on $\alpha$ is contained in the semi-negative definite term $-2 s^i s_i$
involving the norm of the vector $s_i$. This fact allows us to eliminate one redundant direction in
the superpotential parameter space spanned by the $G^i$'s in the analysis of $\sigma$. Indeed, observe 
that $\sigma$ can be maximised with respect to $\alpha$ when $\alpha$ is chosen in such a way that 
$s^i N_i =0$. Since our interest is to determine whether $\sigma>0$ can be achieved, this condition 
fixes $\alpha$ in terms of $N^i$. It also reduces the number of orientations of $G_i$ that need to be analysed 
in order to deduce the convexity of $\sigma$ about the set of stationary points $G_i \propto K_i$. 
Notice additionally that in the particular case of two moduli $i = 1,2$, the condition $s^i N_i =0$ is 
equivalent to $s_i=0$, as there is only one possible direction perpendicular to $K_i$, implying that 
$s_i$ and $N_i$ are parallel to each other.

In the next two sections we study more concretely the function $\sigma$ for the two relevant cases of 
heterotic and orientifold compactifications of string theory.

\section{Heterotic compactifications of string theory} \label{heterotic}
\setcounter{equation}{0}

In this section  we consider a  class of supergravity models which arises in compactifications 
of the heterotic string on Calabi-Yau threefolds.\footnote{Alternatively they can also be viewed 
as the NS-sector of type II compactifications.} Let us first discuss some  generic features of these 
compactifications and then continue with specific examples. 

\subsection{General discussion}

The moduli of heterotic Calabi-Yau compactifications include the dilaton/axion and the deformations 
of the  Calabi-Yau metric. The latter are divided into deformations of the K\"ahler class and deformations 
of the complex structure. Locally, the moduli space $\cal M$ is the product manifold
\be
{\cal M} = {\cal M}^\mathrm{ks}\times {\cal M}^\mathrm{cs}\times \frac{SU(1,1)}{U(1)} \,,
\ee
where ${\cal M}^\mathrm{ks}$ is the space spanned by the K\"ahler moduli, ${\cal M}^\mathrm{cs}$ is spanned 
by the complex structure moduli while the dilaton/axion are the coordinates of the last factor. 
${\cal M}^\mathrm{ks}$ and ${\cal M}^\mathrm{cs}$ are special K\"ahler manifolds in that their K\"ahler
potential can be expressed in terms of a holomorphic prepotential $f=f(\Phi)$. One has 
\cite{Strominger:1985ks, Dixon:1989fj, Candelas:1990pi}
\be\label{specialK}
K = - \log Y, \qquad \textrm{with} \qquad  Y = -2(f+\bff)+(f_k+\bff_{\bar k})(\Phi^k+\bar \Phi^{ k}) \,,
\ee
where in the large-volume limit $Y^\mathrm{cs/ks}$ are given by
\be\label{csks}
Y^\mathrm{cs}=i \int_X \Omega \wedge \bar \Omega \,, \qquad 
Y^\mathrm{ks}= \mathcal{V} \equiv  \frac{4}{3} \int_X J \wedge
J \wedge J  \,.
\ee
Here $\Omega$ and $J$ are,  respectively, the holomorphic $(3,0)$-form and the K\"ahler 
$(1,1)$-form of the Calabi-Yau threefold. $\mathcal{V}$ is the classical volume in that the
equality $Y^\mathrm{ks} = \mathcal{V}$ only holds in the large-volume limit, and it is modified by 
$\alpha'$ and worldsheet-instanton corrections.

There exist various dynamical effects, such as fluxes or gaugino condensates, which can induce 
a nontrivial superpotential $W$ for the moduli \cite{reviews}. We do not systematically discuss 
here all the possible superpotentials but rather assume that most of the moduli are stabilised in a 
supersymmetric way at high energy scales. In addition we assume that supersymmetry is broken by
$F$-terms of the remaining moduli multiplets.\footnote{We similarly assume that matter fields are 
stabilised at supersymmetric points and that their vacuum expectation values remain zero after 
supersymmetry is broken by the moduli.}  This latter sector is the one we want to study in the spirit 
of Sections \ref{sugra} and \ref{S:LVS}. In other words, we want to understand under what conditions 
the moduli sector can simultaneously break supersymmetry and generate a de Sitter vacuum.

For concreteness, let us focus on the K\"ahler moduli sector in the large-volume limit and assume 
that it induces supersymmetry breaking. Of course we could equivalently consider the complex 
structure moduli in the large-complex-structure limit which --due to mirror symmetry--  would lead 
to an identical analysis.

Since $J$ is harmonic, it can be expanded in a $h^{1,1}$-dimensional basis  $w_i, i=1, \ldots,h^{1,1}$ 
of the cohomology group $H^{1,1}$ via $J=v^{i}w_i$. The NS two-form enjoys a similar expansion 
$B_2 = b^i \omega_i$. The coefficients in these expansions $v^i$ and $b^i$ are scalar fields which 
combine into the complex coordinates $T^i=v^i+i b^i$. Inserting this into (\ref{csks}), one obtains
\be
K = - \log \mathcal{V} \ , \qquad \mathrm{with} \qquad \mathcal{V} = 
\frac{1}{6}\, d_{i j k}\, (T^{i} + \bar T^{i})  (T^{j}  + \bar T^{j} )
(T^{k} + \bar T^{k}) \, , \label{eq: K-heterotic}
\ee
where $d_{i j k}=\int_X w_i \wedge w_j \wedge w_k$ are  the Calabi-Yau intersection 
numbers.\footnote{This is indeed a special K\"ahler geometry since $\mathcal{V}$ can be 
derived from the holomorphic prepotential $f(T)=1/6\, d_{i j k} T^i T^j T^k$.}

Before we continue let us emphasise that such a K\"ahler potential also appears as a subsector 
of other string compactifications, for example, in Calabi-Yau compactifications of type IIB with 
$O5/O9$-orientifold planes \cite{Grimm:2004uq}. Therefore the following analysis is not only valid 
for heterotic compactifications but rather for any moduli-sector with a K\"ahler potential of the form 
given in  eq.~(\ref{eq: K-heterotic}).

In order to compute $\sigma$ let us first recall a few further properties of $K$ (for more details on the 
following computations we refer the reader to the appendix). Its first derivative reads
\begin{eqnarray}\label{Kder}
K_{i} = - \frac{\mathcal{V} _{i}}{\mathcal{V} }\ , \qquad
\mathcal{V}_{i}  =  \frac12\, d_{i j k} (T^{j} + \bar T^{j}) (T^{k} + \bar T^{k}) \ .
\end{eqnarray}
The K\"ahler metric is then given by
\begin{eqnarray}\label{metric}
g_{i j}\ =\ - \frac{\mathcal{V}_{ij}}{\mathcal{V}} +
\frac{\mathcal{V}_{i} \mathcal{V}_{j}}{\mathcal{V}^{2}} \
=\ e^K d_{ijk} K^k + K_i K_j 
\ ,  \label{metric-heterotic}
\end{eqnarray}
where the matrix $\mathcal{V}_{ij}= d_{i j k} (T^{k} + \bar T^{k})$ has a signature $(1,h^{1,1}-1)$ for 
all allowed values of $T^{i} + \bar T^{i}$, i.e. those values for which $\mathcal V$ is positive and the 
K\"ahler metric is positive-definite~\cite{Candelas:1990pi}. The inverse metric is conveniently 
expressed in terms of the matrix $\mathcal{V}^{ij}$ which is defined as the inverse of 
$\mathcal{V}_{ij}$, i.e.\  $\mathcal{V}^{ij}\mathcal{V}_{jk} = \delta_k^i$. Using 
$2 \mathcal{V}^{i j} \mathcal{V}_{j} = T^{i}  + \bar T^{i} = - K^i$ one has
\begin{eqnarray}\label{inverse}
g^{ij} = - \mathcal{V} \mathcal{V}^{i j } + \frac{1}{2} K^i K^j.
\end{eqnarray}
From  (\ref{Kder}) and (\ref{inverse}) it follows that $K$ obeys the the no-scale condition (\ref{nosc}) 
and also the homogeneity property (\ref{hom}). 

Using (\ref{Kder})--(\ref{inverse}) one also easily computes the third derivative of $K$ and its 
Riemann tensor: 
\bea
K_{i j k} &=&- e^K d_{ijk} + g_{ij} K_k + g_{ik} K_j + g_{jk} K_i - K_i K_j K_k \,, \label{Kijk}\\
R_{i j m n} &=& g_{i j} g_{m n} + g_{i n} g_{m j} - e^{2K} d_{i m p} g^{p q} d_{q j n} \label{Rijmn} \,. 
\label{curv-no-oplane}
\eea
Notice that the specific form of the Riemann tensor holds for any special K\"ahler manifold with 
$d_{ijk}$ replaced by the third derivative $f_{ijk}$ of the prepotential \cite{de Wit:1984pk, Cremmer:1984hj}.
Inserting (\ref{Rijmn}) into  eq.~(\ref{sigma}) we finally obtain
\begin{eqnarray}\label{S2}
\sigma = - \frac{4}{3} \big(G^i \bar G_i \big)^2 + e^{2K} G^{i} G^{j} d_{i j p} g^{pq} d_{q m n} \bar G^m \bar G^n.
\end{eqnarray}

As in the last section we can rewrite $\sigma$ in terms of $K_i$ and its orthogonal complement $N_i$  
as defined in eqs.~(\ref{decomp}) and (\ref{nialfa}). Inserting (\ref{Kijk}) and (\ref{Rijmn}) into
(\ref{sn}) and (\ref{omega}) we arrive at $\sigma = - 2 s^i s_i + \omega $ with $s_i$ and $\omega$ 
given by
\begin{eqnarray}
\!\! s^i &=& \alpha \bar N^i + \bar \alpha  N^i - \frac 12 e^K P^{ij} d_{j m n} N^m \! \bar N^n \,,\label{s1} \\
\!\! \omega &=& \Big(\!\!-\! \frac 43 g_{ij} \,g_{mn} \!+\! \frac 13 g_{im} \,g_{jn} 
\!+\! \frac{1}{2} e^{2 K} d_{i j p} P^{p q} d_{q m n} \!+\! e^{2 K} d_{i m p} P^{p q} d_{q j n}  \Big) 
N^i \bar N^j N^m \bar N^n \,. \label{w1}
\end{eqnarray}
Let us recall here that with these expressions it is possible now to study the convexity of $\sigma$ by 
scanning $N^i$ and keeping $\alpha$ fixed in such a way that $s^i N_i = 0$.

\subsection{Particular classes of models} \label{examples} 

We now discuss a few specific classes of K\"ahler moduli spaces that can be handled analytically. 
As we shall see, it is possible to obtain examples of models where $\sigma > 0$ for certain directions 
$G^i$ offering the possibility of  generating metastable vacua. Nevertheless, we shall also see that 
there are entire classes of models for which $\sigma$ is unavoidably negative-definite, implying the
existence of at least one tachyonic state in the spectrum which renders the theory
unstable independently of the form of the superpotential. 

\subsubsection{Factorisable K\"ahler manifolds}

As our first example we discuss Calabi-Yau threefolds which are $K3$-fibrations over a ${\bf P_1}$-base. 
In the limit of a large ${\bf P_1}$ the K\"ahler potential simplifies and reads \cite{Klemm:1995tj, Aspinwall:1995vk}
\bea
K = - \log \Big(\frac 12\, d_{1ab} (T^1 + \bar T^1) (T^a + \bar T^a) (T^b + \bar T^b) + \dots \Big) \,, 
\label{khet}
\eea
where $T^{1}$ parametrises the volume of the ${\bf P_1}$-base while the $T^a, a=2,\dots, h^{1,1}$ are 
moduli of the $K3$ fibre. The dots indicate further cubic terms which, however, are independent of $T^1$ 
and therefore subleading in the large ${\bf P_1}$-limit. In that limit the K\"ahler metric is block diagonal 
($g_{1 a} = 0$) and hence the moduli space factorises into the special K\"ahler space\footnote{This
also uses the fact that the matrix $d_{1ab}$ has signature $(1,h^{1,1}-2)$.}
\be \label{SO}
{\cal M}^\mathrm{ks} \ =\ \frac{SU(1,1)}{U(1)} \times
\frac{SO(2,h^{1,1}-1)}{SO(2)\times SO(h^{1,1}-1)}\ .
\ee
The K\"ahler potential also enjoys the properties 
\be\label{nsspec}
K^1K_1=1\ ,\qquad 
K^a K_a=2\ .
\ee

In order to compute $\sigma$ we observe that (\ref{metric}) implies  
$d_{1 a b} = e^{-K} K_1\,\big(g_{a b} - K_a K_b\big)$ which, together with 
(\ref{nsspec}), leads to
\bea
e^{2K}\, d_{1 a c}\, d^c{}_{1 b} = g_{1 1}\, g_{a b}\ ,  \qquad
e^{2K} d_{a b 1}\, d^1{}_{c  e} =  \big(g_{a b} - K_a K_b\big)\,\big(g_{ c e} - K_{c} K_{ e} \big) \,.
\eea
Inserting this into (\ref{S2}) we obtain
\bea
\sigma = -\frac 43 (G^1\bar G_1 + G^a \bar G_a)^2 + |G_a G^a - (K_a G^a)^2|^2 
+ 4\, \big(G^1 \bar G_1 \big)\big(G^a \bar G_a \big) \,.
\label{sigmahet1}
\eea

To find an upper bound for this function, we use the inequality $|A \cdot B |^{2} \leq |A|^2 |B|^2$ for
$A_{a} = (g_{ab} - K_a K_b)\, G^b$ and $B_{a} =  G_{a}$. This together with (\ref{nsspec}) yields
\begin{eqnarray}
|G_a G^a - (K_{a} G^{a})^{2}|^{2} \le  (G^a \bar G_a)^{2} \,. \label{ineq-K3}
\end{eqnarray}
As a consequence, the function $\sigma$ given in  eq.~(\ref{sigmahet1}) obeys
\bea
\sigma \le - \frac13 \, \big(2 \, G^1 \bar G_1- G^a \bar G_a \big)^2 \,.
\eea
We see that $\sigma$ is always negative and vanishes along the flat direction where 
$2 G^1 \bar G_1 = G^a \bar G_a$. This means that the preferred supersymmetry breaking direction 
is $G^i \propto K^i$ as for models with constant curvature. We conclude that in this class of models 
one always has a tachyonic sGoldstino, which can at best become massless for Minkowski vacua and 
for a special Goldstino direction.

Note that the scalar manifold (\ref{SO}) associated with these factorisable models is a constant curvature 
coset manifold. The implications of the metastability condition for this type of models were also studied in
ref.~\cite{GomezReino:2006wv}. It was in particular shown that the second factor 
in (\ref{SO}) behaves effectively as two copies of the first factor, independently of $h^{1,1}$. 
This implies that the metastability condition for $K3$ fibrations is analogous to that of models with 
$3$ independent moduli, as in eq.~(\ref{K-fact}) with $n_i=1$, providing an alternative derivation
of the fact that $\sigma$ is at best zero in these models.

\subsubsection{Two-field models} \label{diagonal inters}

Another class of models that can be studied analytically are those with only $2$ moduli 
$T^i=v^i+ib^i$, with $i=1,2$. To perform this analysis we recall that $\sigma$ may be 
written as $\sigma =  - 2 s^i s_i + \omega$ with $s_i$ and  $\omega$ given by eqs.~(\ref{s1}) and 
(\ref{w1}) respectively. In the case of $2$ moduli it was shown in Section \ref{S:homogeneous no-scale} 
that it is always possible to choose $s_i = 0$, thereby maximising $\sigma$. We are thus left with 
the task of computing the function $\omega$ and check if $\omega>0$ is allowed. As can be read 
from (\ref{w1}), the function $\omega$ depends on the variables $N_i$. Since these are orthogonal 
to $K^i$, they can be parameterised with a single complex quantity $C$ as
\be \label{param}
(N^1,N^2) = (K_2,-K_1)\, C \,.
\ee
With this definition, one has $N^i N_i = 3 \det g \, |C|^2$.

One first case that we can analyse is the case of models with only diagonal intersection numbers 
$d_{111}$ and $d_{222}$. In this example the K\"ahler potential takes the form
\be\label{p1}
K = - \log \Big( \frac 16 d_{111} (T^{1} + \bar T^{1})^3 + \frac 16 d_{222} (T^{2} + \bar T^{2})^3 \Big) \,.
\label{Khet1}
\ee
Computing the metric and its inverse, and using eqs.~(\ref{param}) and (\ref{Khet1}) with (\ref{w1}), 
we find that
\begin{eqnarray}\label{s0}
\omega = \frac {81}{8} \, e^{4 K} \, \frac {d_{111}^2 d_{222}^2}{\det g} \, |C|^4\,. \label{omega>0}
\end{eqnarray}
This result is positive since the metric has to be positive definite. 
This shows that $\sigma$ can be made positive and that the stability condition 
can be fulfilled for certain particular directions of $G_i$. As shown for general large-volume scenarios, 
we find that the point $N^i=0$, where $G^i\propto K^i$, is indeed a 
stationary point with $\sigma=0$. Nevertheless, as can be read off from (\ref{s0}), 
in this case this stationary point is a saddle point, and $\sigma(G^i)$ can actually be made 
positive along some directions.

By now we have shown that in the case of factorisable K\"ahler potentials we get $\omega=0$ and 
in the case of diagonal intersection numbers we get $\omega>0$. But one may wonder whether 
in some cases one can have $\omega < 0$. In order to answer this question, 
let us consider a model with the following K\"ahler potential:
\be
K = - \log \Big(\frac 12 \, d_{122} \, (T^1 + \bar T^{1}) (T^2 + \bar T^{2})^2 
+ \frac 16 \, d_{111} (T^1 + \bar T^{1})^3\Big)\,.
\ee
Note now that in the limit $d_{111} \to 0$ this K\"ahler potential becomes of the form 
(\ref{khet}) describing factorisable models, for which the maximal value of $\sigma$ is zero. 
One can then study how this result is modified in the case where $d_{111} \ll d_{122}$
by performing an expansion in the small parameter
\be
\epsilon = \frac {d_{111}}{d_{122}} \,.
\ee 
Following now the same strategy as before it is straightforward to find that
\begin{eqnarray}\label{om2}
\omega = \frac {81}{2} \, \epsilon \, e^{4 K} \, \frac {d_{122}^4}{\det g} \, |C|^4\,.
\end{eqnarray}
This result can be either positive or negative depending on the sign of $\epsilon$. This implies
that $\sigma$ can be positive or must be negative, depending on the sign of $\epsilon$. 

Actually, for these two-field models it is possible to compute the function $\omega$ for generic 
values of all the independent intersection numbers $d_{111}$, $d_{222}$, $d_{122}$ and $d_{112}$. 
Using the general form for the K\"ahler potential (\ref{eq: K-heterotic}) and following the same steps as 
in the previous examples one finds, after some algebra, that the value of $\omega$ can be cast into 
the simple form
\be\label{om3}
\omega = - \frac {3}{8} \, e^{4 K} \, \frac {\Delta}{\det g} \, |C|^4\,,
\ee
where the quantity $\Delta$ is the discriminant of the cubic polynomial defined by $d_{ijk} v^i v^j v^k$ 
after scaling out one variable, and reads
\be\label{del}
\Delta = - 27 \Big( d_{111}^2 d_{222}^2 - 3\, d_{112}^2 d_{122}^2 + 4\, d_{111} d_{122}^3 + 4\, d_{112}^3 d_{222} 
- 6\, d_{111} d_{112} d_{122} d_{222} \Big)\,. 
\ee
Since we must require $\det g > 0$, the sign of $\omega$ is fixed by the sign of $\Delta$.
Moreover, it becomes now clear that the two categories of models with $\omega > 0$ and $\omega<0$ 
are of comparable size and that they merge in the very special class of models with factorisable K\"ahler 
geometries, for which $\omega=0$. 

\subsection{Including $\alpha'$ corrections}   \label{corrections} 

So far we have analysed models respecting the no-scale property  $K_i K^i = 3$. This property is however
violated when $\alpha'$, worldsheet instanton or string loop corrections to the K\"ahler potential are 
taken into account, although they are suppressed in the large-volume and weak-coupling limit. It is therefore 
interesting to study how the bounds on the mass of the sGoldstinos are modified by these effects, particularly 
for those models in which $\sigma \le 0$  at leading order. For concreteness we here consider only 
$\alpha'$ corrections, but the effect of  other corrections can be studied in a similar way.

When $\alpha'$ corrections are taken into account, the K\"ahler potential is 
$K = -  \log Y$ where~\cite{Candelas:1990rm}
\bea
Y  =  \mathcal{V} + 4 \xi \,.
\eea
The quantity $\xi=-\zeta(3) \chi/2$ is a real constant determined by the Euler characteristic of the 
Calabi-Yau manifold, given by $\chi=2(h^{1,1}-h^{2,1})$. The geometry is still of the special-K\"ahler
type, with prepotential $f(T) = 1/6\, d_{i j k} T^{i} T^{j} T^{k} - \xi$. However, as mentioned above, $\alpha'$ corrections 
break the no-scale property (\ref{nosc}), which is seen from eqs.~(\ref{eq:theta}) and (\ref{eq:generalK})
of the appendix with $n=1$ and $\theta = (3/2) \mathcal{V}/(\mathcal{V} + 4 \xi)$.

The natural small dimensionless parameter controlling the effect of $\alpha'$ corrections 
relative to the leading-order K\"ahler potential is given by
\be
\delta = \frac {4 \xi}{\mathcal{V}} \,.
\ee
In the following, we work at leading order in this parameter, which is small when the volume 
is large. Using eqs.~(\ref{eq:Ki}) and (\ref{eq:generalK}) with $\theta \simeq 3/2(1 - \delta)$, one then finds that
\be
K_i K^i \simeq 3 + 6\, \delta \,.
\ee
The Riemann tensor is given by eq.~(\ref{eq:RijmnSK}). The quantities $f_{i j k}$ are as before given 
by the intersection numbers, whereas the metric $g_{i j}$ and its inverse $g^{i j}$ are affected by the 
corrections and can be computed from~(\ref{eq:Ki}). 

In order to understand how the corrections modify the bounds on the sGoldstino masses, it is useful 
to compute the function $\sigma(G^i)$ up to second order in the $N^i$'s and at leading order in $\delta$. 
One finds
\bea \label{eq:sigmahetalpha}
\sigma(G^i) &\simeq& 120\, \delta\, |\alpha|^4 - 4\, (1- 2\, \delta) \, |\alpha|^2 g_{i j}N^i \bar N^{j} \nn \\[1mm]
&\;& -\, 2\, (1+9\, \delta) \, \Big(\alpha^2 g_{i j} \bar N^i \bar N^j + {\rm c.c.} \Big) + \mathcal O(N^3) \,. 
\label{eq:sigmamax}
\eea
Notice that $\sigma$ continues to be stationary at $N^i = 0$, but its value at that point becomes 
$\sigma_0 \simeq 120 \, \delta \, |\alpha|^4$. If $\chi<0$ (i.e. $h^{2,1}>h^{1,1}$)  then this is 
positive and the special direction $G^i \propto K^i$ always allows to fulfil the metastability condition.

Up to this point we have left $\alpha$ undetermined. We can however express $|\alpha|^2$ in terms of 
the vacuum energy density $V= e^G(G^i G_i -3)$ and  gravitino mass scale $m_{3/2} = e^{G/2}$ as
$|\alpha|^2 = 1 + V/(3\, m_{3/2}^2) + \mathcal O(\delta)$. Inserting this relation back into eq.~(\ref{eq:sigmamax}) 
and evaluating at $N^i = 0$ one obtains
\be \label{eq:sigmastathet}
\sigma_0 \simeq 120\, \delta\, \Bigg( 1 + \frac{V}{3\, m_{3/2}^2} \Bigg)^2 \,. \label{eq:newsigma-alpha}
\ee
This relation can be used to compute the mass scale $\tilde m^{2} = e^{G} \lambda / G^{i} G_{i}$, as introduced 
in Section  \ref{stability analysis}, at the critical value $G^i \propto K^i$. This is particularly important for models
where $\sigma \le 0$ at leading order, as it then provides a bound on the attainable values of the sGoldstino mass. 
By inserting eq.~(\ref{eq:newsigma-alpha}) into eq.~(\ref{lambda}), and specialising to the relevant regime 
$V/m_{3/2}^2 \ll 1$, one obtains
\be
\frac{\tilde m^{2}}{m_{3/2}^{2}} \simeq 40\, \delta - \frac 2 3  \frac{V}{m_{3/2}^2} \,.
\ee
It immediately follows that if $\delta \gtrsim V / (60\, m_{3/2}^2 )$ then the metastability condition is fulfilled. 
This gives a criterion on how large $\alpha'$ corrections have to be for given gravitino scale and vacuum energy 
density in order to admit viable vacua. Notice that under these circumstances the value of the sGoldstino mass 
is essentially the gravitino mass suppressed by $\alpha'$ corrections. We should bear in mind, however, that 
other corrections to the K\"ahler potential could compete against $\alpha'$ corrections and modify this result.

\section{Orientifold compactifications of string theory} \label{orientifolds}
\setcounter{equation}{0}

\subsection{General discussion}

In contrast to the heterotic string, type IIB  Calabi-Yau compactifications give theories with $\mathcal N=2$ 
supersymmetry in 4 dimensions. The RR forms which are present in 10-D type II supergravities lead to additional 
massless 4-D fields which, together with the geometric moduli, arrange into $\mathcal N=2$ supermultiplets. 
The scalars in the vector multiplets span again a special K\"ahler manifold ${\cal M}^\mathrm{SK}$ whereas the scalars 
in the hypermultiplet span a dual quaternionic manifold ${\cal M}^\mathrm{Q}$. 

One way to obtain a theory with $\mathcal N=1$ supersymmetry is to impose an orientifold projection. In type IIA, 
this involves $O6$-planes while in type IIB one has $O3/O7$ or $O5/O9$-planes. The moduli space in all of these 
three cases has the form \cite{Grimm:2004uq, Grimm:2004ua, Grimm:2005fa}
\be
\tilde{\cal M} = \tilde{\cal M}^\mathrm{SK} \times \tilde{\cal M}^\mathrm{Q}\ ,
\ee
where $\tilde{\cal M}^\mathrm{SK}$ is a special K\"ahler submanifold of the ``parent'' $\mathcal N=2$ moduli space 
${\cal M}^\mathrm{SK}$ while $\tilde{\cal M}^\mathrm{Q}$ is a K\"ahler submanifold of ${\cal M}^\mathrm{Q}$. In the 
large-volume large-complex-structure limit, the $\tilde{\cal M}^\mathrm{SK}$~factor satisfies the no-scale property 
and the K\"ahler potential does in fact coincide with the K\"ahler potential of eq.~(\ref{eq: K-heterotic}). 
Therefore the analysis of Section~\ref{heterotic} holds unmodified for the moduli of $\tilde{\cal M}^\mathrm{SK}$. 
On the other hand the $\tilde{\cal M}^\mathrm{Q}$~sector, which includes the dilaton, satisfies $K^i K_i =4$, and 
if the dilaton is fixed, the latter sector is also no-scale~\cite{Grimm:2004uq}. However, the K\"ahler potential of 
$\tilde {\cal M}^{Q}$ is different for the three orientifold compactifications. 

For concreteness let us focus on type IIB with O3/O7 planes, where the K\"ahler potential in the large-volume 
limit reads \cite{Grimm:2004uq}
\be \label{eq:orientifoldkaehler}
K_Q= -2 \log \mathcal{V} - \log (S + \bar S) \,, \qquad \mbox{with}\qquad  \mathcal V =\frac{1}{48}\, d^{i j k}
v_{i} v_{j} v_{k} \ .
\ee
$\mathcal V$ is again the classical volume of the Calabi-Yau orientifold, $S$ is the dilaton/axion and the 
$v_{i}, i = 1, \ldots , h_+^{1,1}$ are the K\"ahler moduli of the Calabi-Yau orientifold. However the $v_i$ do 
not appear as components of chiral multiplets in the low energy effective action. Instead, they determine the 
real part of the K\"ahler coordinates $T^{i} = \rho^{i} + i \zeta^{i}$ via the quadratic relation\footnote{Strictly 
speaking there can also be $h_{-}^{1,1}$ moduli $G$ with couplings specified in~\cite{Grimm:2004uq} which 
however we neglect during the analysis of this paper.}
\be
\rho^i=\frac{1}{16}\, d^{i j k} v_{j} v_{k}\ . \label{eq: var-change}
\ee
Due to this relation the K\"ahler potential of eq.~(\ref{eq:orientifoldkaehler}) cannot explicitly  be expressed in 
terms of the coordinates $T^{i}$, but is only implicitly defined through eq.~(\ref{eq: var-change}).\footnote{In 
order to comply with the standard notation whereby chiral coordinates carry upper indices, we have slightly 
abused the notation by lowering the indices of $v$ and raising them for the intersection numbers $d$. We have also
rescaled the intersection numbers as $d_{i j k} \rightarrow d^{i j k} / 8$.
However we stress that they are exactly the same objects as in the heterotic case.} As in the previous section 
we assume that the dilaton is fixed to a supersymmetric configuration and focus only on the K\"ahler moduli.

The metric can be conveniently expressed in terms of 
\be\label{ddef}
d^{ij} \equiv \frac {\partial \rho^i}{\partial v_j} = \frac{1}{8}\,
d^{i j k} v_{k}\ , \qquad 
d_{ij} \equiv \frac {\partial v_i}{\partial \rho^j}    \ .
\ee
Using (\ref{eq:orientifoldkaehler}) -- (\ref{ddef}), one computes
\begin{eqnarray}\label{KOder}
K_{i} = -\,\frac 12\, e^{K/2} v_i \ , \qquad 
d^{ij} = - \, \frac 14 \, e^{-K/2} d^{ijk} K_k \ .
\end{eqnarray}
This in turn determines the K\"ahler metric and its inverse to be
\begin{eqnarray}
g_{i j} = \frac{1}{2} \, K_{i} K_{j} - \frac 14 \, e^{K/2} d_{ij} \, ,  \qquad 
g^{i j} = 4 \, \rho^i \rho^j - 4 \, e^{-K/2} d^{ij}\,.  \label{inv-K 1} 
\end{eqnarray}
One can now check that $K$ satisfies the no-scale property $K^i K_i = 3$ as well as the special 
identity $K^{i} = - 2  \rho^i$, which again results from the fact that $e^{-K}$
is a homogeneous function of degree $3$ in $\rho^i$. This can be used
to slightly rewrite the inverse metric as
\be\label{inv-K 4}
g^{i j} =  e^{-K} d^{i j k} K_{k} + K^{i} K^{j}\ .  \label{inverse-orientifolds}
\ee
Notice that this expression for the inverse metric is equal in form to the metric (\ref{metric-heterotic}) 
of the heterotic case.  Similarly, the inverse metric of the heterotic case is equal in form to the metric 
(\ref{inv-K 1}) for the orientifold case examined here. As was shown in ref.~\cite{D'Auria:2004kx} this 
property directly follows from the fact that in the orientifold case the K\"ahler coordinates $T^i$ feature 
the dual variables $\rho^i$ instead of $v_i$ as the real part.

In order to determine $\sigma$ we need again the third derivatives of
the K\"ahler potential and the Riemann tensor. For this it is convenient to
first compute derivatives of $g^{i j}$. Using the above relations we find
\bea
\left[g^{i j} \right]_{k} &=& e^{-K}  d^{i j m} g_{m k} 
- (g^{i j} - K^{i} K^{j}) K_{k} - \delta^{i}_{k} K^{j} -
\delta^{j}_{k} K^{i} \,,  
\nn\\[1mm]
\left[ g^{i j} \right]_{m n}  &=& - e^{- 2K}   d^{i j p} g_{p q} 
d^{q r s}  g_{r m} g_{s n} + \delta^{i}_{m} \delta^{j}_{n}
+ \delta^{i}_{n}\delta^{j}_{m} \,.  \label{two-der}
\eea
$K_{ijm}$  and the Riemann tensor are expressed in terms of these
derivatives as
\bea
K_{ijm} &=& - g_{i p} [g^{p q}]_j g_{q m} \,, \nn\\ 
R_{i j m n} &=&  - g_{i p} g_{q j} [g^{p q}]_{m n} +  g_{i r} [g^{r
   p}]_m g_{p q} [g^{q s}]_n g_{s j} \, .
\label{Riemann-orientifold}
\eea
Inserting (\ref{two-der}) into (\ref{Riemann-orientifold}) and using
(\ref{ddef})--(\ref{inv-K 1}) we arrive at\footnote{This Riemann tensor was also computed in 
ref.~\cite{D'Auria:2004cu}.}
\bea\label{Rijmno}
K_{ijm} &=& e^{-K} \dm_{ijm} - g_{i j} K_m  - g_{im} K_j - g_{jm} K_i
+ K_i K_j K_{k} \ , 
\nn\\[1mm]
R_{i j m n} &=& - g_{im} g_{jn} + e^{-2K} \big(\dm_{i j k} g^{kl} \dm_{l m n} 
+ \dm_{i n k} g^{kl} \dm_{l j m} \big)  + g_{in} K_j K_m + g_{jm} K_i K_n  \nn\\
&\;&+\, g_{im} K_j K_n + g_{jn} K_i K_m + g_{ij} K_m K_n + g_{mn} K_i K_j - 3 K_i K_j K_m K_n \nn \\
&\;& -\, e^{-K} \big(\dm_{imj} K_n + \dm_{imn} K_j + \dm_{inj} K_m +
\dm_{nmj} K_i \big)  \ ,
\eea
where we abbreviated
\be
\dm_{ijk} \equiv g_{ip} g_{jq} g_{kl} d^{p q l} \ . 
\ee
Inserting (\ref{Rijmno}) into (\ref{sigma}) we finally arrive after
some algebra at
\begin{eqnarray}
\sigma &=&  \frac{2}{3} \big(G^i \bar G_i \big)^{2}  + |G^{i} G_{i} - (K^{i} G_{i})^{2} |^{2} + 2 |K^{i} 
G_{i}|^{4} - 4|K^{i} G_{i}|^{2}  G^j \bar G_j  \nn\\
&&- 2\, e^{-2K}  G_{i} \bar G_{j} d^{i j p} g_{p q} d^{m n q} G_{m} \bar G_{n} 
+ 2\, e^{-K} d^{i j k} G_{i} \bar G_{j}  (G_{k} K^{n} \bar G_{n} + \bar G_{k} K^{n} G_{ n}) \,. \quad 
\label{sigma-string}
\end{eqnarray}

It is also possible to write $\sigma$ in terms of the decomposition $G_{i} = N_{i} + \alpha K_{i}$ defined in 
(\ref{decomp}). Doing so, one finds the result (\ref{sigmaalphaN}) or (\ref{sigmaomegas}), with the quantities 
$g_{ij}$, $K_{ijk}$ and $R_{ijmn}$ given by eq.~(\ref{Rijmno}). Again only the few terms transverse 
to $K^i$ contribute in contractions with $N^i$. The quantities $s_i$ and $\omega$ are obtained by inserting
(\ref{Rijmno}) into (\ref{sn}) and (\ref{omega}) and are given by
\begin{eqnarray}
\omega &=&  \Big(g^{im} g^{jn} - \frac 32 \,e^{-2 K}  d^{i j p} P_{p q} d^{p m n} \Big)
N_{i}  \bar N_{j} N_{m} \bar N_{n} \,,\label{w2} \\
s_{i}  &=&  \alpha N_{\bar \imath} + \bar \alpha  N_{i} - \frac 12 e^{-K} P_{ij} d^{j m n} N_{m} N_{\bar n} \,.
\label{s2}
\end{eqnarray}
It is interesting to compare both of these quantities with their heterotic counterparts, given in eqs.~(\ref{s1})  and 
(\ref{w1}). While $s_i$ of eq.~(\ref{s1}) is equal in form to the one given here, $\omega$ of eq.~(\ref{w1}) has
essentially the opposite sign to the one shown here, and involves the inverse metric instead of the metric 
and $e^{-K}$ instead of $e^K$.
As we shall see, this result is particularly relevant for models with two moduli, for which $\sigma$ can be maximised
by setting $s_i=0$ and thus the sign of $\sigma$ is determined by $\omega$.

\subsection{Particular classes of models}

As for  the heterotic case, we can only make further progress by
computing $\sigma$ for specific classes of Calabi-Yau orientifolds.
In the following we consider the same examples as in Section \ref{examples}.

\subsubsection{Factorisable K\"ahler manifolds}

We again start with $K3$-fibred Calabi-Yau threefolds where the K\"ahler potential 
takes the form 
\bea\label{KK3}
K = - 2\,\log \Big(\frac 1{16}\, d^{1ab}  v_1 v_a  v_b \Big) \,.
\eea
For these intersection numbers $(v_1,v_a)$ can be explicitly determined in terms of the 
$(\rho^1,\rho^a)$ via (\ref{eq:  var-change}). One finds 
$v_1 = 2 \, (d_{1ab}^{-1} \rho^a \rho^b/\rho^1)^{1/2}$ and 
$v_a = 4 \, d_{1ab}^{-1} \rho^b (\rho^1/ d_{1cd}^{-1} \rho^c \rho^d)^{1/2}$.
Inserting into (\ref{KK3}) and using $\rho^{i} = (T^i + \bar T^i)/2$ yields
\be
K = - \log \Big[\frac{1}{2} \, d_{1ab}^{-1} (T^1 + \bar T^1) (T^a + \bar T^a) (T^b + \bar T^b) \Big] \,.
\ee
This is exactly the same $K$ as in the heterotic case but with an inverse intersection matrix. 
In particular, $K$ obeys again $K^1K_1=1$ and $K^aK_a=2$. It is nevertheless instructive to 
recompute the function $\sigma$ by using the formulae obtained for orientifold models.
{}From eq.~(\ref{inv-K 4}) we first infer $d^{1 a b} = e^{K} K^{1} (g^{a b} - K^{a} K^{b})$. 
This allows us to compute
\begin{eqnarray}
e^{-2K} G_{i} \bar G_j d^{i j p} d_p^{m n} G_m  \bar G_n 
&=& (G^a \bar G_{a} - |K^{a} G_{a}|^{2})^{2} + G^a G_a (K^{1} \bar G_1)^{2} \nonumber\\ 
&\;& +\, \bar G^a \bar G_a (K^{1}  G_{1})^{2} + 2\, G^{a} G_{a} |K^{1} G_{1}|^{2} \,, \\
e^{-K} d^{i j k} G_{i}  \bar G_j G_{k} K^{n} \bar G_n 
&=& 2 \,(K^{1} \bar G_{1} + K^{a} G_{\bar a}) K^{1} G_{1} ( G^{b} G_{b} - |K^{b} G_{b}|^{2}) \nonumber\\ 
&\;& +\, (K^{1} \bar G_{1} + K^{a} \bar G_{a}) K^{1} \bar G_{1} (G^{b} G_{b} - (K^{b} G_{b})^{2}) \,.
\end{eqnarray}
Inserting into (\ref{sigma-string}) we arrive at
\begin{eqnarray}
\sigma &=& - \frac{1}{3} (G^a \bar G_a + |K^{1} G_{1}|^{2} )^{2}
+ |G^{a} G_{a} - (K^{a} G_{a})^{2}|^{2} -  (G^a \bar G_{a}  - |K^{1} G_{1}|^{2} )^{2} \,.
\end{eqnarray}
Using the same inequality (\ref{ineq-K3}) as for heterotic models, and noticing also the simplification 
$|K^1 G_1|^2 = G^1 \bar G_1 $, one finally deduces the same upper bound as before:
\begin{eqnarray}
\sigma \le - \frac{1}{3}  (2 \, G^1 \bar G_1 - G^a \bar G_a)^{2} \,.
\end{eqnarray}
Therefore, we arrive at the same conclusion as for heterotic models: in this class of factorisable models 
the stability bound is always at least marginally violated.

\subsubsection{Two-field models} 

As for heterotic models, another class of models where the analysis simplifies are those involving 
two fields. In such a situation, there is again a single direction $N^i$ orthogonal to $K_i$, which 
can be parametrised as
\be \label{param2}
(N_1,N_2) = (K^2,-K^1) C \,.
\ee
With this definition, one has $N^i N_i = 3 / \det g \,|C|^2$.
As before using this parametrisation we can compute the value of the quantity $\omega$ defined 
by (\ref{w2}), which provides an upper bound to $\sigma$.

As for the heterotic models we consider first the simplest case of models with only diagonal 
intersection numbers $d^{111}$ and $d^{222}$. The corresponding K\"ahler potential is of the 
form\footnote{In ref.~\cite{D'Auria:2004cu} the same manifold was studied as an example where 
the Riemann tensor of the manifold and its dual manifold do not coincide.}
\be\label{KOf}
K = - 2\, \log \Big( \frac 1{48} d^{111} v_1^3 +  \frac 1{48} d^{222} v_2^3  \Big)\,.
\ee
Using (\ref{eq:  var-change}) one determines $v_1 = 4\, (\rho^1/d^{111})^{1/2}$ and 
$v_1 = - 4\, (\rho^2/d^{222})^{1/2}$ which, when inserted back into (\ref{KOf}), yields
\begin{eqnarray}
K = - 2 \log \Big( \frac {\sqrt{2}}3 (d^{111})^{-1/2} (T^1 + \bar T^1)^{3/2} 
- \frac {\sqrt{2}}3 (d^{222})^{-1/2} (T^2 + \bar T^2)^{3/2} \Big) \,.
\end{eqnarray}
The function $\omega$ is now easily computed and is found to be
\begin{eqnarray}\label{Omor0}
\omega = - \frac {81}{8} \, e^{- 4 K} \,(d^{111})^2 (d^{222})^2\, \det g \, |C|^4 \,.
\end{eqnarray}
This result is negative and shows that in this case $\sigma \le 0$ for any choice of $G_{i}$. It is therefore 
impossible to obtain stable de Sitter vacua in this case.  Furthermore, this inequality is saturated only for 
$N_{i} = 0$, which corresponds to the configuration $G_{i} \propto K_{i}$. 
The result presented here should be contrasted to the one presented in eq.~(\ref{omega>0}).

To understand whether this negative sign for $\omega$ persists or not in more general $2$-field models, 
let us as before consider a small deformation of a factorisable model. The simplest example has 
non-zero $d^{122}$ and $d^{111}$, and a K\"ahler potential 
given by
\be
K = - 2 \log \Big(\frac 1{16}\, d^{122} v_1 v_2^2 + \frac 1{48}\, d^{111} v_1^3 \Big)\,.
\ee
In the limit $d^{111} \ll d^{122}$, in which the model is nearly factorisable, one can expand 
at leading order in the small parameter
\be
\epsilon = \frac {d^{111}}{d^{122}} \,.
\ee 
One finds $v_1 = 2\, (d^{122} \rho^1)^{-1/2} \rho^2 [1 + 
\epsilon/8 (\rho^2/\rho^1)^2]$ and $v_2 = 4\, (d^{122}/\rho^1)^{-1/2} [1 - \epsilon/8 (\rho^2/\rho^1)^2]$. 
The K\"ahler potential can then  be rewritten as
\begin{eqnarray}
K = - \log \Bigg(\frac 12  \frac 1{d^{122}} (T^1 + \bar T^1) (T^2 + \bar T^2)^{2} 
- \frac 1{24} \frac {d^{111}}{(d^{122})^2} \frac {(T^2 + \bar T^2)^4}{T^1 + \bar T^1}\Bigg)\,.
\end{eqnarray}
After a straightforward computation, the function $\omega$ is found to be
\begin{eqnarray}\label{Omor}
\omega = - \frac {81}{2} \, \epsilon \, e^{-4 K} \, (d^{122})^4 \, \det g \, |C|^4\,.
\end{eqnarray}
As in the heterotic case we have again that this result can be either positive or negative, 
depending on the sign of $\epsilon$. This means that also in orientifold compactifications 
one can have models with $\sigma > 0$ and models with $\sigma <0$.

Note that the results (\ref{Omor0}) and (\ref{Omor}) take the same form as
(\ref{s0}) and (\ref{om2}) for heterotic models but with the substitutions
$e^K \to e^{- K}$, $\det g \to (\det g)^{-1}$ and a flip in the overall sign.
This is due to the fact that in the case of two-field models, where the
parametrisations (\ref{param}) and (\ref{param2}) can be used, the functions
(\ref{w1}) and (\ref{w2}) get indeed precisely mapped into each other by these
substitutions. This map can then be used to infer that also for orientifold models
the result for generic intersection numbers $d^{111}$, $d^{222}$, $d^{122}$ and
$d^{112}$ should take a simple form, obtained by applying it to the heterotic
result (\ref{om3}). This leads to the result
\be
\omega = \frac {3}{8} \, e^{- 4 K} \, \Delta \,\det g \, |C|^4\,,
\ee
in terms of the discriminant
\bea
\Delta &=& -27 \Big((d^{111})^2 (d^{222})^2 - 3\, (d^{112})^2 (d^{122})^2 + 4\,d^{111}(d^{122})^3  \nn \\
&\;& \hspace{26pt} +\, 4\, d^{222} (d^{112})^3 - 6\, d^{111} d^{112} d^{122} d^{222} \Big) \,.
\eea
It is not straightforward to verify this result explicitly, because performing the change
of variables (\ref{eq: var-change}) involves in this general case finding the roots of a quartic
polynomial. But we were nevertheless able to verify it by brute force with computer assistance.
Since we must require $\det g > 0$, the sign of $\omega$ is again determined by the sign of the
quantity $\Delta$, which has exactly the same structure as for heterotic models.

It is important to note that the results found for heterotic and orientifold models imply 
that for any given string compactification with non-zero $\Delta$, one can have either viable 
heterotic models but no viable orientifold models (if $\Delta <0$), or vice-versa (if $\Delta > 0$).

\subsection{Including $\alpha'$ corrections}

We now include $\alpha'$ corrections in orientifold compactifications.
When these corrections are taken into account, the K\"ahler
potential of eq.~(\ref{eq:orientifoldkaehler}) is modified to $K = - 2
\log Y - \log (S + \bar S)$, where~\cite{Becker:2002nn}
\be
Y = \mathcal{V} + \frac{\xi}{2} \, \left(  \frac{S + \bar S}{2} \right) ^{3/2}. \label{Y-alpha}
\ee
One difficulty arises from the fact that these corrections depend on the dilaton which, strictly speaking, now should be considered as a dynamical quantity (this is due to the fact that in the presence of $\alpha'$ corrections the K\"ahler potential is not factorisable). 
To simplify the presentation of this section we nevertheless assume that the dilaton can be fixed to a
constant value in eq.~(\ref{Y-alpha}), and define the new constant 
$\tilde \xi = (\xi/2) [(S + \bar S)/2]^{3/2}$.\footnote{Similar conclusions are obtained in the full computation 
with a dynamical dilaton by assuming that $S$ is fixed to a supersymmetric configuration $G_{S} = 0$.}
As before, $\alpha'$ corrections break the no-scale property (\ref{nosc}), which can be seen from 
eqs.~(\ref{eq:theta}) and (\ref{eq:generalK}) of the appendix with $n=2$ and 
$\theta = 3 \mathcal{V}/(\mathcal{V} + \tilde \xi)$.

The small dimensionless parameter controlling the relative effect of the $\alpha'$ corrections is in this
case given by
\be
\tilde \delta = \frac {\tilde \xi}{8 \mathcal{V}} \,.
\ee
We will work at leading order in this parameter. Using the results of the 
appendix with $\theta \simeq 3(1 - 8 \tilde \delta)$, one finds then that
\be
K_i K^i \simeq 3 + 12 \, \tilde \delta \,.
\ee
The Riemann tensor, given by eq.~(\ref{eq:generalRiemann}), can be evaluated by using 
$Y_{i j} = 1/8\, d_{i j}$, $Y_{i j m} = - 1/128\, d_{i r} d_{j s} d_{m t} d^{r s t}$ and 
$Y_{i j m n} = 24\, Y_{i j s} d^{s r} Y_{r m n}$.

As worked out in Section \ref{corrections} for the case of heterotic compactifications, one may compute
$\sigma$ up to second order in $N^{i}$ and at first order in $\tilde \delta$:\footnote{For this computation, 
the following contractions are needed: 
$Y_{i j} K^i K^j = 3 \, (\theta -1)^{-2} \mV$,
$Y_{i j m} K^i K^j = Y/2 \, (\theta -1)^{-2} K_m$, 
$Y_{i j m n} K^i K^j K^m K^n = 9\,(\theta -1)^{-4} \mV$.}
\bea 
\sigma(G^i) &\simeq& 105\, \tilde \delta\, |\alpha|^4 - 4\, (1+ 14\, \tilde \delta) \, |\alpha|^2 g_{i j}N^i \bar N^{j} \nn \\[1mm]
&\;& -\, 2\, (1+ 27\, \tilde \delta) \, \Big(\alpha^2 g_{i j} \bar N^i \bar N^j + {\rm c.c.} \Big) + \mathcal O(N^3) \,. 
\eea
Again, $\sigma$ is stationary at $N^i = 0$ with a value $\sigma \simeq  105\, \tilde\delta\, |\alpha|^{4}$. 
Observe that the only difference with respect to the result found for heterotic models, shown in 
eq.~(\ref{eq:sigmamax}), is the numerical factor in front of $\tilde \delta$. 
We can now calculate the mass scale $\tilde m^{2} = e^{G} \lambda / G^{i} G_{i}$ associated to the sGoldstino.
By repeating the steps of Section~\ref{corrections} and assuming that $V / m_{3/2}^{2} \ll 1 $ one 
arrives at
\be
\frac{\tilde m^{2}}{m_{3/2}^{2}} \simeq 35 \, \tilde \delta - \frac 2 3 \frac{V}{m_{3/2}^{2}} \,.
\ee
Similarly to the case of heterotic compactifications, if $\tilde \delta \gtrsim 2 V / (105 m_{3/2}^{2})$ then the 
metastability condition is fulfilled and the sGoldstino mass becomes of the order of the gravitino mass
suppressed by $\alpha'$ corrections. This is for instance the case in the models of ref.~\cite{Balasubramanian:2005zx,Balasubramanian:2004uy}.

\section{Conclusions} \label{conclusions} 
\setcounter{equation}{0}

In this paper we have analysed the role that neutral chiral multiplets have in the construction of 4-D metastable vacua, 
paying special attention to the generic class of models obtained in large-volume compactifications of string theory. 
In general, metastable vacua with spontaneously broken supersymmetry are only granted in models where a 
non-vanishing $F$-term $F^{i}= m_{3/2} G^{i}$ exists such that $\sigma(G^i) > 0$, as defined in eq.~(\ref{sigma}). 
This necessary condition was shown to be equivalent to the requirement of having a positive square mass for the 
sGoldstinos when the vacuum energy density $V$ is non-negative. Interestingly, this condition was also shown 
to be sufficient, with the understanding that all of the other scalar fields can be given arbitrarily large positive 
square masses if the superpotential of the theory is suitably tuned. 

In the particular case of  large-volume string compactifications the function $\sigma$ respects some severe restrictions.
For instance, from the general analysis made in Section \ref{S:LVS}, we have learned that the set of values $G_i \propto K_i$ corresponds to a family of stationary points of $\sigma$ with $\sigma = 0$. Moreover, they are either saddle points 
or maxima, depending on the intersection numbers of the particular model. 
Despite of the difficulties posed by a complete analytical study of the function $\sigma$ we were still able to outline 
a general procedure to determine whether a particular compactification admits dS vacua. This procedure was 
introduced first for generic supergravity models in Section  \ref{S:analysis} and then refined in Section 
\ref{S:homogeneous no-scale} for the particular case of string compactifications. We believe that such a
procedure can be implemented numerically and should be of considerable help in any computer scan of 
string ground states. We also saw, however, that there are interesting and nontrivial examples of compactifications 
which can be handled analytically. For $K3$ fibrations, for instance, we showed that $\sigma$ can be at best 
zero. For $2$-field models, on the other hand, the maximal value of $\sigma$ can be non-vanishing, and its 
sign is controlled by the discriminant $\Delta$ of the cubic polynomial defined by the intersection numbers.
Moreover, for $\Delta < 0$ one can find viable heterotic models but no viable orientifold model, and vice-versa
for $\Delta > 0$.

The results of this paper are useful for determining which type of configurations within a given model should help 
in the construction of vacua. We have seen for example that exploring configurations in the superpotential 
parameter space close to the critical point $G^i \propto K^i$ give a vanishing value for $\sigma$ and that 
$\alpha'$ corrections can help in obtaining a positive --although suppressed-- square mass for the sGoldstinos, 
independently of whether $\sigma>0$ is admitted or not at leading order. In fact, one could expect this to be a 
generic feature of any additional sector which breaks the no-scale property $K^i K_i = 3$ respected by the K\"ahler 
moduli sector.

Finally, let us mention here that a strategy similar to the one used in this paper could be used also to study the 
possibility of constructing successful models of slow-roll inflation within a string-theoretical scenario. This requires 
finding some direction in field space with small first and second derivatives of the potential. The first condition 
corresponds approximately to stationarity, whereas the second one requires a small negative mass. The algebraic 
problem defined by these two conditions is then very similar to the one faced in this paper \cite{workprog}. 

\section*{Acknowledgements}

This work was partly supported by the German Science Foundation (DFG)
under the Collaborative Research Center (SFB) 676, by the European Union 
6th Framework Program MRTN-CT-503369 ``Quest for
unification" and by the Swiss National Science Foundation. 
LC would like to thank the Institute of Theoretical Physics
of the Warsaw University for their hospitality during the final
stages of this work. LC is supported by a Maria Curie Transfer
of Knowledge Fellowship of the European Community's Sixth Framework
Programme under contract number MTKD-CT-2005-029466 (2006-2010).
CS thanks the University of Hamburg and DESY for their hospitality during the 
beginning of this work. 

\begin{appendix}

\section{Details of K\"ahler geometries} \label{Appendix}   

In this appendix, we collect some useful formulae concerning the geometry of K\"ahler and special-K\"ahler
manifolds, which are needed in some derivations in the main text.

\renewcommand{\theequation}{\thesection.\arabic{equation}}
\setcounter{equation}{0}  

\subsection{Logarithmic K\"ahler potentials} \label{sec:generalformulae}

Let us consider a K\"ahler potential of the form $K= - n \log Y$, where $Y$ is some real function 
of the scalar fields $\phi^i$ and $n$ is a real number. Denoting by $Y^{i \bj}$ the inverse of 
$Y_{i \bj}$, one easily finds
\bea
K_i &=& - n \frac {Y_i}{ Y} \,, \label{eq:Ki} \nn \\
g_{i \bj} &=& -  n \frac{Y_{i \bj}}{Y}+ n \frac{Y_i Y_{\bj}}{Y^2} = -  n \frac{Y_{i \bj}}{Y}+ \frac 1 n K_{i} K_{\bj} \,, \nn \\
g^{i \bj} &=& - \frac{Y Y^{i \bj}}{n} + \frac 1 n \frac{1}{\theta - 1} Y^{i \br} Y_{\br} Y^{\bj s} Y_s = - \frac{Y Y^{i \bj}}{n} 
+ \frac{\theta -1}{n} K^{i} K^{\bj} \,, \nn \\
K^i &=&- \frac{1}{\theta - 1}Y^{i \br} Y_{\br} \,.
\eea
The quantity $\theta$ is defined as
\be \label{eq:theta}
\theta \equiv \frac{Y_i Y^{i \bj} Y_{\bj}}{Y} \,,
\ee
and controls the value of the contraction defining the no-scale property:
\bea 
K^i K_i &=& n \frac{\theta}{\theta-1} \,. \label{eq:generalK}
\eea
The third derivatives of $K$ are 
\bea
K_{i \bj m} &=& - \frac{n}{Y}  Y_{i \bj m}+ \frac{n}{Y^{2}}  (Y_{i} Y_{\bj m} \! + Y_{m} Y_{\bj i} +  Y_{\bj}  Y_{i m} )
- \frac{2n}{Y^3}  Y_{i} Y_{\bj} Y_{m} \,, \nn \\
K_{i \bj \bn} &=&  - \frac{n}{Y} Y_{i \bj \bn}+ \frac{n}{Y^{2}} (  Y_{\bj} Y_{i \bn} \! + Y_{\bn} Y_{i \bj} +  Y_{i}  Y_{\bj \bar n}  )
- \frac{2n}{Y^3}  Y_{i} Y_{\bj} Y_{\bar n} \,.
\eea
Finally, the Riemann tensor for the K\"ahler manifold is
\bea \label{eq:generalRiemann}
R_{i \bj m \bn} &=& K_{i \bj m \bn} - K_{i m \br} g^{\br s} K_{s \bj \bn} \nn \\
&=& \frac 1 n \big(g_{i \bj} \, g_{m \bn} + g_{i \bn} \, g_{m \bj}\big) 
- \frac n Y Y_{i \bj m \bn}  - \frac{n}{Y^2} ( n Y_{i m \bs} g^{\bs r}Y_{r \bj \bn} 
+ \frac{1}{\theta -1} Y_{im} Y_{\bj \bn} ) \nn \\
&\;& + \ \frac{n^2}{Y^3} ( Y_{i m} Y_{\bj \bn r} g^{r \bs} Y_{\bs} + Y_{\bj \bn} Y_{i m \bs} g^{\bs r} Y_{r} ).
\eea

\subsection{Special K\"ahler manifolds}

We now consider the case of special K\"ahler geometries, for which the K\"ahler potential
$K = - \log Y$ itself admits a holomorphic prepotential $f$, in terms of which
\be
Y = -2(f+\bff)+(f_k+\bff_{\bar k})(\phi^k+\bar \phi^{ k}).
\ee
The Riemann tensor simplifies substantially in this case. Indeed, one easily computes
$Y_i + Y_{\bar \imath} = N_{i j} \big(\phi^j + \bar \phi^{\bar \jmath} \big)$ and $Y_{i \bar \jmath} = N_{ij}$, 
where $N_{ij} = f_{ij} + \bar f_{\bar \imath \bar \jmath}$. Combining these two expressions, one gets 
then $Y^{i \bar \jmath} (Y_j + Y_{\bar \jmath}) = (\phi^i + \bar \phi^{\bar \imath})$. Finally, combining this result 
with $Y_{i j} = f_{ijk} \big(\phi^k + \bar \phi^{\bar k} \big)$ and $Y_{i j \bar k} = f_{ijk}$, one 
obtains the relation $Y_{i j \bar s} Y^{\bar s r} \big(Y_r + Y_{\bar r}\big) = Y_{ip}$. Using these relations,
one finally finds~\cite{Cremmer:1984hj}
\be \label{eq:RijmnSK}
R_{i \bj m \bn} = g_{i \bj} \, g_{m \bn} + g_{i \bn} \, g_{m \bj} - \frac{1}{Y^2}f_{i m r}g^{r \bs}\bff_{\bs \bj \bn}.
\ee

\end{appendix}

\end{document}